**Title:**

A Generalized Muon Trajectory Estimation Algorithm with Energy Loss for Application to Muon Tomography

**Authors:**

Stylianos Chatzidakis,[*] Zhengzhi Liu,[†] Jason P. Hayward,[†,*] and John M. Scaglione[*]
[*]Oak Ridge National Laboratory, Oak Ridge, TN  37831
[†]Department of Nuclear Engineering, University of Tennessee, Knoxville, TN  37996

**Contact Info:**

Tel: +1 865-576-8981, Email: chatzidakiss@ornl.gov

**Abstract:**

This work presents a generalized muon trajectory estimation (GMTE) algorithm to estimate the path of a muon in either uniform or nonuniform media. The use of cosmic ray muons in nuclear nonproliferation and safeguards verification applications has recently gained attention due to the non-intrusive and passive nature of the inspection, penetrating capabilities, as well as recent advances in detectors that measure position and direction of the individual muons before and after traversing the imaged object. However, muon image reconstruction techniques are limited in resolution due to low muon flux and the effects of multiple Coulomb scattering (MCS). Current reconstruction algorithms, e.g., point of closest approach (PoCA) or straight-line path (SLP), rely on overly simple assumptions for muon path estimation through the imaged object. For robust muon tomography, efficient and flexible physics-based algorithms are needed to model the MCS process and accurately estimate the most probable trajectory of a muon as it traverses an object. In the present work, the use of a Bayesian framework and a Gaussian approximation of MCS are explored for estimation of the most likely path of a cosmic ray muon traversing uniform or nonuniform media and undergoing MCS. The algorithm's precision is compared to Monte Carlo simulated muon trajectories. It was found that the algorithm is expected to be able to predict muon tracks to less than 1.5 mm RMS for 0.5 GeV muons and 0.25 mm RMS for 3 GeV muons, a 50% improvement compared to SLP and 15% improvement when compared to PoCA. Further, a 30% increase in useful muon flux was observed relative to PoCA. Muon track prediction improved for higher muon energies or smaller penetration depth where energy loss is not significant. The effect of energy loss due to ionization is investigated, and a linear energy loss relation that is easy to use is proposed.

# 1. Introduction

Cosmic ray muon tomography was first proposed and explored experimentally by Borozdin et al. in 2003 [1], and a number of advantages of muon tomography over conventional imaging modalities, e.g., x-rays, were identified. Recently, the use of muons in nuclear nonproliferation and safeguards verification applications has received attention as a passive and penetrating way to interrogate objects. This endeavor has been assisted by the development of detectors that can measure and provide information related to the incoming and outgoing trajectories of individual muons [2–6]. Recently, cosmic ray muons have been investigated for volcano imaging [7–9] and cargo scanning applications [10–12]. Their use has been extended to nuclear waste imaging [13–16] and determination of nuclear fuel debris location in nuclear reactors having suffered from the effects of a severe accident such as the one that occurred in Fukushima [17, 18].

Some of the challenges of muon tomography are low muon flux at 10,000 muons/s/cm$^2$, difficulty in muon momentum measurement due to high energy (>1 GeV) and velocity leading to low energy loss per unit length, and the substandard resolution of the available imaging reconstruction algorithms [19-24]. In addition, recent efforts exploring muon computed tomography (μCT) are limited by the tendency of muons to scatter in the target, thus blurring the image. No direct information about the muon path traversing the medium under interrogation is available, and some type of extrapolation is required for muon imaging. Current reconstruction algorithms rely on simple assumptions for muon path estimation through the imaged object. One common assumption is to use a straight-line path defined by the line between the intersection of the entry and exit path lines [22, 23]. Another commonly used assumption is based on the calculation of the point of closest approach (PoCA) between the incoming and outgoing trajectories [20, 21]. The algorithm calculates the shortest perpendicular distance between incoming and outgoing trajectories and assigns the scattering event to a voxel located at the middle of that distance. Besides the crude approximations, muon trajectories may not intersect, so some muon events must be discarded, thereby increasing the measurement time to achieve an acceptable image quality. Another major limitation is the allocation of points outside of the region of interest. These points are rejected, reducing the useful muon flux and increasing the measurement time requirements.

These disadvantages could be partially alleviated by estimating the path of a muon when it traverses an object, along with a probability envelope. This work presents a physics-based algorithm coupled with Bayesian theory, a Gaussian approximation of MCS, and generalized scattering and displacement moments. The proposed algorithm is motivated by a prior formalism introduced by Shulte et al. [25] which was developed to calculate the most probable trajectory of protons while traversing a uniform material. However, this approach was not extended to include high-energy cosmic ray muons or nonuniform high-Z materials. Furthermore, it required the evaluation of complicated ratios of polynomials to account for energy loss effects. In contrast, the algorithm described in this work employs a bivariate Gaussian approximation of MCS with the generalized scattering and displacement moments to estimate the path of a muon in either uniform or nonuniform media. The energy loss is calculated using the continuous slowing down approximation and exploits the minimum ionization property of muons to derive a linear relation that is easy to use. The estimated muon path asymptotically approaches the incoming and outgoing muon trajectories. The algorithm was tested under various scenarios using detailed Geant4 Monte Carlo [15, 21, 26] simulations. It is hypothesized that the proposed generalized muon trajectory estimation (GMTE) algorithm will produce muon tomographic images with improved quality and reduced noise, requiring fewer muons than the muon reconstruction techniques that are currently available.

## 2. Cosmic Ray Muon Tomography

Existing detectors for muon tomography measure the locations and directions of incoming and outgoing muons. The detectors are typically parallel planes of position sensitive chambers, such as scintillators [2] drift-wire chambers [3, 4] or gas-electron multiplier detectors, although cylindrical detectors have also been proposed [22]. A typical configuration of some detectors and an imaged object in muon tomography applications is shown in Figure 1. Multiple planes are needed to record the direction of each muon. In addition, each muon will result in four or more recorded position measurements. Existing detectors employ several X-Y layers of strips of scintillation fibers [2] or drift-wire chambers [3] arranged in orthogonal rows, achieving moderate to high resolution (in certain cases ~50 microns). Only muons that pass through all detector planes are recorded. This information is then processed using muon tomographic algorithms that estimate the muon trajectory and reconstruct the object under interrogation. Only muons trajectories that fall into the superimposed reconstruction grid boundaries are useful for reconstruction. With this knowledge, low, medium, and high Z materials can be discerned, and density maps can be reconstructed with centimeter spatial resolution using iterative reconstruction algorithms.

The discrimination capability and image quality are usually limited by muon flux and incomplete knowledge about muon path and muon momentum [27]. Optimal measurement time for muon tomographic applications is a balance between two trade-offs. Using fewer muons provides significant time savings but lower quality images, while a higher number of measured muons results in improved image contrast and resolution but increased measurement time. Existing studies on muon-CT detectors have demonstrated that muons on the order of millions (requiring minutes to days of measurement time, depending on the size and orientation of the detectors) are required for imaging large scale objects such as cargo containers [4, 5, 11] or dry casks [21-23]. Recent work based on simulations showed that several days (>$10^6$ muons) are needed to identify the location of spent nuclear fuel in dry casks [21–23]. It is noted that the simulated results did not include any additional measurement time needed to address noise and/or detector measurement uncertainties. An experimental effort by Durham et al. [28] resulted in ~200 hours of measurement time to register $1.62 \times 10^5$ muons. Given the size and orientation of the detectors, this is approximately 100 times lower than theoretically possible. Any additional improvement in image efficiency, i.e., muons used for imaging to muons measured, will allow for imaging using fewer muons, thus reducing time measurement requirements.

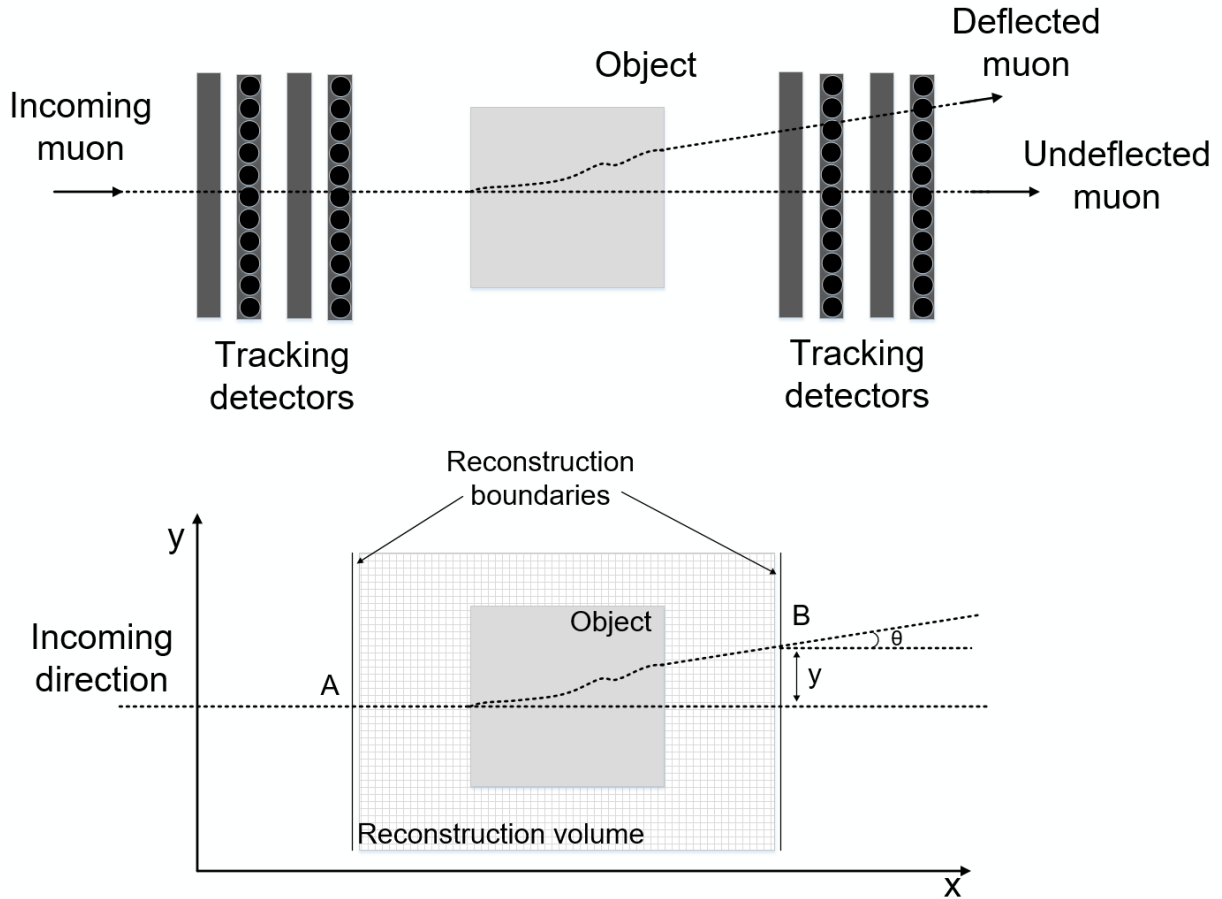

Figure 1. Two-dimensional view of a typical configuration of some detectors and an imaged object in muon tomography applications (top). Tracking detectors measuring incoming and outgoing trajectories of individual muons; reconstruction geometry used throughout this work (bottom).

3. Generalized Muon Trajectory Estimator

In this work, we sought to replace said assumptions by a physics-based model in an effort to improve the fidelity of the reconstructions. To quantify how likely it would be for a muon to have scattered to a particular scattering angle θ and displacement y (see Fig. 1), a maximum-a-posteriori algorithm is used that maximizes the posterior distribution derived from Bayes' theorem. The scattering and displacement probability distributions are based on a Gaussian approximation of the physics-derived distributions that describe the MCS processes. Generalized scattering and displacement moments are used to describe the distribution parameters. MCS is a random process that tends to change the directions of muons while they traverse a target. MCS involves many individual elastic interactions between muon and nuclei without changing the muon momentum. Several theories that describe MCS have been formulated to characterize the scattering probability of a particle traversing an object at a certain depth. Of these theories, Moliere's theory [29], which remains analytical to the end and was later improved by Bethe [30], is in good agreement with most of experimental data [31-34]. Although Moliere's derived scattering distribution is generally complex, a Gaussian approximation has been shown to adequately represent the central 95% of the scattering distribution [31]. In 3D space, a muon can be deflected in two perpendicular planes that are uncorrelated and can thus be treated independently. As such, a bivariate (one parameter for scattering angle and one for displacement) Gaussian approximation is assumed in what follows. The equations presented in this paper can be used twice in a muon image reconstruction program to account for scattering in two dimensions. Let **Y** be a vector that contains the displacement $y$, and scattering angle $\theta$, of a muon as it traverses an object. The Gaussian approximation of the bivariate scattering distribution is:

$$\mathcal{P}(Y) \sim \mathcal{N}(0, \Sigma), \quad (1)$$

where **Σ** is the covariance matrix:

$$\Sigma = \begin{bmatrix} \sigma_y^2 & \sigma_{\theta y}^2 \\ \sigma_{\theta y}^2 & \sigma_\theta^2 \end{bmatrix}, \quad (2)$$

where $\sigma_y^2$ is the scattering displacement variance, $\sigma_\theta^2$ is the scattering angle variance and $\sigma_{\theta y}^2$ is the covariance between the displacement and the scattering angle. The values of these quantities can be estimated from simulation, theory, or from actual muon measurements. Let **y**$_A$ be a vector at the entry point A and **y**$_j$ be a vector at a point j. The vectors **y**$_A$ and **y**$_j$ contain the displacement $y$ and angle $\theta$ at each point. The probability that a muon will have displacement and angle **y**$_j$ given the exit point is B ($j$ is a point between A and B) is:

$$P(y_j|y_B) = \frac{P(y_B|y_j)P(y_j|y_A)}{P(y_B)}. \quad (3)$$

The left term is the posterior distribution that we seek to maximize. The nominator contains the likelihood probability distribution and prior probability. The probability distribution in the denominator can be considered as a normalization term:

$$P(y_j|y_B) = \frac{1}{C_0} P(y_B|y_j)P(y_j|y_A), \quad (4)$$

where $C_0 = P(y_B)$ is a normalization constant that falls out during maximization. The Gaussian approximation of the bivariate scattering distribution at a point $j$ within the material given **y**$_A$ is

$$P(y_j|y_A) = \frac{1}{C_1} \exp\left(-\frac{1}{2}(y_j^T - y_A^T R_A^T)\Sigma_j^{-1}(y_j - R_A y_A)\right), \quad (5)$$

where $C_1$ is again a normalization constant, and:

$$\Sigma_j = \begin{bmatrix} \sigma_{yj}^2 & \sigma_{\theta yj}^2 \\ \sigma_{\theta yj}^2 & \sigma_{\theta j}^2 \end{bmatrix}, \quad (6)$$

$$R_A = \begin{bmatrix} 1 & z_j - z_A \\ 0 & 1 \end{bmatrix}, \quad (7)$$

where $\Sigma_j$ is the co-variance at point $j$. Similarly, the Gaussian approximation of the bivariate scattering distribution at exit point B within the material given $y_j$

$$P(y_B|y_j) = \frac{1}{C_2}\exp\left(-\frac{1}{2}(y_B^T - y_j^T R_B^T)\Sigma_B^{-1}(y_B - R_B y_j)\right), \tag{8}$$

where $C_2$ is a normalization constant and:

$$\Sigma_B = \begin{bmatrix} \sigma_{yB}^2 & \sigma_{\theta yB}^2 \\ \sigma_{\theta yB}^2 & \sigma_{\theta B}^2 \end{bmatrix}, \tag{9}$$

$$R_B = \begin{bmatrix} 1 & z_B - z_j \\ 0 & 1 \end{bmatrix}, \tag{10}$$

Where $\Sigma_B$ is the covariance at exit point B. Substituting the probability distributions (5) and (8) in Eq. (4) and taking the logarithm

$$lnP(y_j|y_B) = -\frac{1}{2}(y_j^T - y_A^T R_A^T)\Sigma_j^{-1}(y_j - R_A y_A) - \frac{1}{2}(y_B^T - y_j^T R_B^T)\Sigma_B^{-1}(y_B - R_B y_j) - lnC_3, \tag{11}$$

where $C_3$ contains all the other constants $C_0$, $C_1$, and $C_2$. Differentiating with respect to $y_j$ and setting it to zero, the muon path formula can be obtained:

$$y_{MLP} = \left(\Sigma_j^{-1} + R_A^T \Sigma_B^{-1} R_A\right)^{-1}\left(\Sigma_j^{-1} R_A y_A + R_A^T \Sigma_B^{-1} y_B\right). \tag{12}$$

For a muon entering the reconstruction volume parallel to the x-axis and with no lateral displacement, the equation simplifies to:

$$y_{MLP} = \left(\Sigma_j^{-1} + R_A^T \Sigma_B^{-1} R_A\right)^{-1}\left(R_A^T \Sigma_B^{-1} y_B\right). \tag{13}$$

The $2\sigma$ and $3\sigma$ envelopes can be calculated by taking the variance in displacement at depth $j$ from the first row and the first column of the error matrix.

### 4. Muon Energy Loss

Equations (12) and (13) require calculation of the covariance matrices which contain the scattering and displacement moments. Different treatments with various degrees of complexity and accuracy have been published on the multiple scattering of charged particles by matter [32–34]. Due to relativistic effects, the scattering displacement moments are inversely proportional to the muon momentum. The complete calculation takes into account the loss of momentum of a muon as it traverses two points in a given volume, from $x_0$ to $x_1$ [25]:

$$\sigma_\theta^2 = A_0 \left[1 + 0.038 \ln\left(\int_{x_0}^{x_1}\frac{x}{X_0(x)}dx\right)\right]^2 \int_{x_0}^{x_1}\frac{1}{\beta(x)^2 p(x)^2}\frac{dx}{X_0(x)} \quad (rad^2), \tag{14}$$

$$\sigma_y^2 = A_0 \left[1 + 0.038\ln\left(\int_{x_0}^{x_1} \frac{x}{X_0(x)} dx\right)\right]^2 \int_{x_0}^{x_1} \frac{(x_1 - x)^2}{\beta(x)^2 p(x)^2} \frac{dx}{X_0(x)} \ (cm^2), and \qquad (15)$$

$$\sigma_{\theta y}^2 = A_0 \left[1 + 0.038\ln\left(\int_{x_0}^{x_1} \frac{x}{X_0(x)} dx\right)\right]^2 \int_{x_0}^{x_1} \frac{x_1 - x}{\beta(x)^2 p(x)^2} \frac{dx}{X_0(x)} \ (rad \cdot cm), \qquad (16)$$

where $X_0$ is the radiation length (cm), $p$ is muon momentum (MeV), and $\beta$ is the ratio of particle's speed to the speed of light. The radiation length $X_0$, $\beta$, and $p$ are a function of distance and material density. For high energy muons traversing low density, small sized objects, this dependence is often neglected because the relative loss of energy is less than a few percent. However, corrections may be necessary for large objects containing high-Z materials such as uranium. For instance, a 3 GeV muon will lose approximately 0.5 GeV, or 16% of its initial energy when traversing a uranium cube having 20 cm dimension.

To develop an analytical expression for energy loss, let the factor $1/\beta^2 p^2$ in the above expressions be rewritten as a function of the muon kinetic energy T and the traversed depth $x$. First, the momentum becomes

$$p = \frac{1}{c}\sqrt{E_{tot}^2 - E_{rest}^2} = \frac{1}{c}\sqrt{T^2 + 2TE_{rest}}, \qquad (17)$$

where $E_{tot}$ and $E_{rest}$ are the muon total energy and the muon rest mass, respectively. In addition, the ratio of the muon speed to the speed of light $\beta$ is

$$\beta = \left(1 + \frac{E_{rest}^2}{(pc)^2}\right)^{-0.5}. \qquad (18)$$

Combining the above expressions, it can be shown that:

$$\frac{1}{\beta(x)^2 p(x)^2} = \frac{(T(x) + E_{rest})^2}{(T(x) + 2E_{rest})^2 T(x)^2}. \qquad (19)$$

The spatial dependence of Eq. (19) can be derived from Monte Carlo simulations by fitting to simulated data. For example, it has been proposed to model the energy loss using a fifth-degree polynomial [25]:

$$\frac{1}{\beta(x)^2 p(x)^2} = a_0 + a_1 x + a_2 x^2 + a_3 x^3 + a_4 x^4 + a_5 x^5. \qquad (20)$$

However, a simpler functional form can be derived by applying the Bethe-Bloch formula [30] which provides a more complete picture of the energy loss mechanisms including corrections for relativistic and nonparticipation of the strongly bound inner shell electrons. The Bethe-Bloch formula is

$$-\frac{dE}{dx} = 4\pi N_A r_e^2 m_e c^2 Z_1^2 \frac{Z_2}{A_2} \frac{1}{\beta^2} \left[ \frac{1}{2} \ln \frac{2 m_e c^2 \beta^2 \gamma^2 T_{max}}{I^2} - \beta^2 - \frac{\delta}{2} \right] \; (MeV/cm), \quad (21)$$

where $N_A$, $r_e$, $m_e$, $Z_1$, $Z_2$, $A_2$ are Avogadro's number, Bohr radius, electron mass, incoming particle charge, medium atomic number and mass number, respectively. The quantity $T_{max}$ is the maximum energy loss in a single collision with an electron:

$$T_{max} = \frac{2 m_e c^2 \beta^2 \gamma^2}{1 + 2\gamma m_e/M + (m_e/M)^2}. \quad (22)$$

For the energy range relevant to cosmic ray muons, i.e., 1 GeV<E<50 GeV and mean energy 3-4 GeV, and most materials, radiative effects account for less than 1% and nuclear loss rate is negligible. For practical purposes, energy loss can be assumed constant:

$$\frac{dE}{dx} = -a. \quad (23)$$

For muon energies in the range 1 GeV<E<50 GeV and in most materials, the energy loss is 1-2 MeV/cm²g. The constant α ranges from ~2 to ~20 MeV/cm, with water having the lowest value of 1.992 MeV/cm (losses in air or other gases are negligible) and uranium 20.5 MeV/cm. Since the cosmic ray muon energy is much larger than the muon rest mass (105.7 MeV/c²), a reasonable approximation is E=pc, resulting in

$$p(x) = p_0 - ax. \quad (24)$$

With this result, equation (19) can be approximated as:

$$\frac{1}{\beta(x)^2 p(x)^2} \cong \frac{1}{\beta(x)^2 (p_0 - ax)^2} \cong \frac{1}{(p_0 - ax)^2}, \quad (25)$$

where we have explicitly used the relativistic limit $\beta(x) \cong 1$. Taking the integral, we obtain:

$$\int \frac{dx}{\beta(x)^2 p(x)^2} = \int \frac{dx}{p(x)^2} = \int \frac{dx}{(p_0 - ax)^2} = \frac{x}{p_0(p_0 - ax)} + C. \quad (26)$$

In order to explore whether Eq. 26 captures the salient characteristics for the limits of the problem the energy loss behavior of the momentum factor $1/\beta^2 p^2$ as a function of penetration depth was simulated using Geant4. The average value of energy loss as a function of traversed depth for three initial muon energies (3, 5, and 10 GeV) traversing 80 cm of a high-Z object (uranium, Z=92, density 18.95 g/cm³, α=20.5 MeV/cm) is shown in Figure 2. For the considered three muon energies, the energy loss is approximately linear and the gradient is 20.05 MeV/cm at 3 GeV, 20.58 MeV/cm at 5 GeV, and 21.20 MeV/cm at 10 GeV. This justifies the assumption of a constant energy loss per unit length. The average value of the quantity $1/\beta^2 p^2$ was also investigated, and the results from the Geant4 simulations were compared with the constant energy loss approximation. The results, shown in Figure 3 and summarized in Table 1, are in good agreement within a few per cent with higher differences observed at lower energies where energy losses are more pronounced.

Table 1. Geant4 energy loss results and comparison with linear energy loss approximation (Eq. 25).

| | Geant4 | | | Eq. 25 approximation | |
|---|---|---|---|---|---|
| Initial Energy | Final energy | Percent energy loss | Energy loss, dE/dx | Constant α | Mean absolute percentage difference from Geant4 |
| 3000 MeV | 1390.25 MeV | 53.65% | 20.05 MeV/cm | 20.50 MeV/cm | 10.20% |
| 5000 MeV | 3349.43 MeV | 33.02% | 20.58 MeV/cm | 20.50 MeV/cm | 5.90% |
| 10000 MeV | 8301.25 MeV | 16.98% | 21.20 MeV/cm | 20.50 MeV/cm | 2.29% |

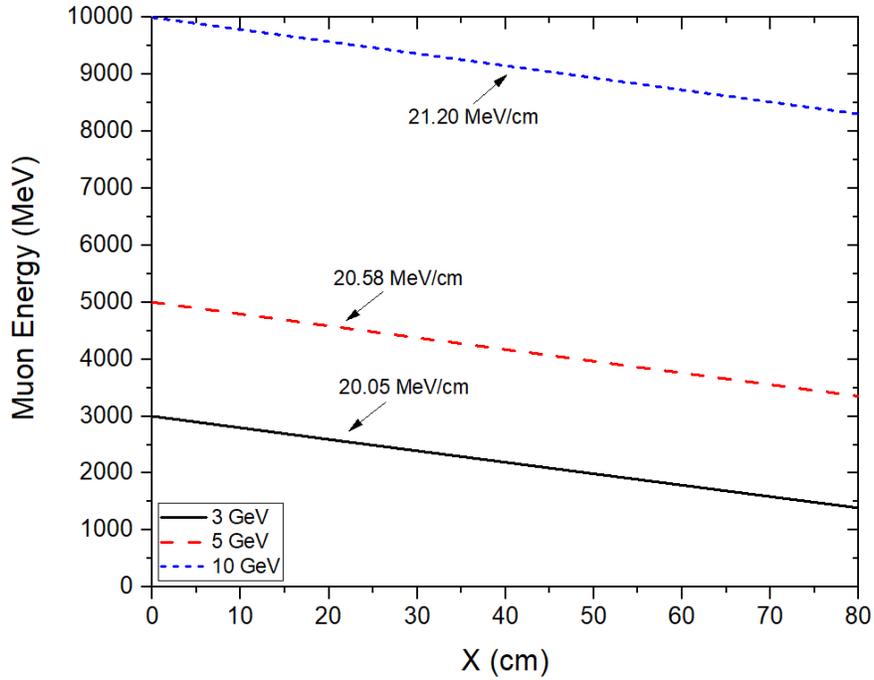

Figure 2. Geant4 simulated muon energy as a function of traversed depth for uranium and three initial muon energies (3, 5, and 10 GeV). The energy loss is linear in all cases, showing a constant energy loss per unit length behavior.

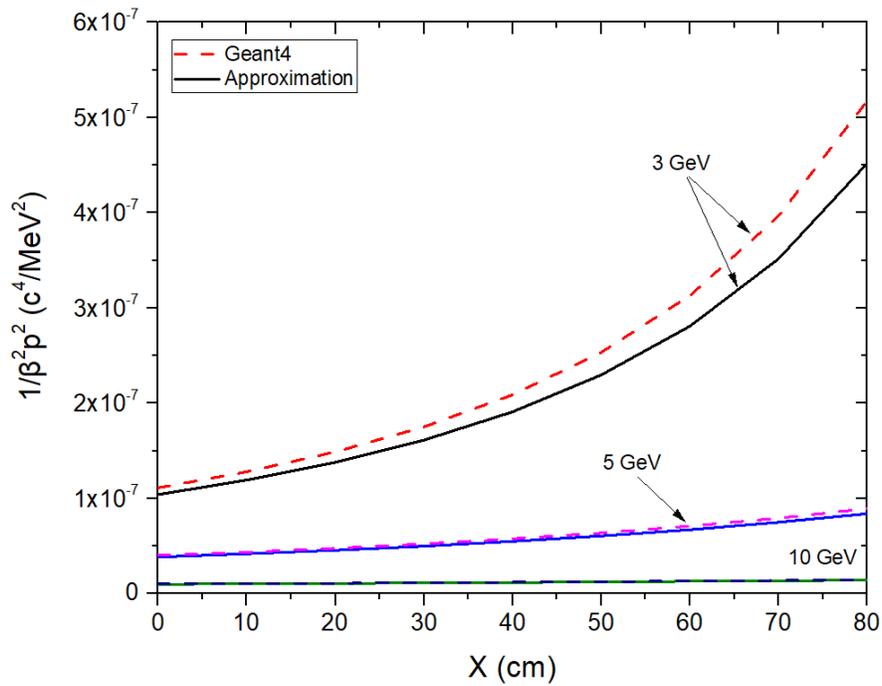

Figure 3. Average value of $1/\beta^2 p^2$ as a function of penetration depth in uranium for three initial muon energies (3, 5 and 10 GeV).

## 5. Geant4 Simulations

To simulate MCS, Geant4 implements a non-Gaussian model that is validated with experimental data [15, 21], and it is used in the paper as a tool to test the proposed GMTE algorithm. Due to the interest in using cosmic ray muons in spent nuclear fuel storage cask assay, uranium represents high-Z material, iron represents medium-Z material, and concrete represents low-Z material. Two different cases were tested using Geant4: (a) a uniform cube with a depth of 10 cm, and (b) a uniform cube with a depth of 80 cm. The large uniform cube with a depth of 80 cm was selected to study the effect of energy loss. For case (a), muons with energies 1, 3, 5, or 10 GeV were simulated. For case (b), 3 GeV muons were simulated. The cases are summarized in Table 2.

Table 2. Geant4 test scenarios.

| Test case | Muon energy (GeV) | Material | Size (cm) |
|---|---|---|---|
| (a) | 1, 3, 5, 10 | U, Fe, Concrete | 10 |
| (b) | 3 | U | 80 |

Figures 4 and 5 demonstrate examples of muon tracks in a uniform material with depths of 10 or 80 cm and the associated GMTE envelopes. The straight-line path (SLP) and PoCA lines are also shown. In all cases, the estimated GMTE is in good agreement with the simulated muon path. Note that there are cases in which the PoCA path is outside the reconstruction boundaries. These trajectories will normally be rejected during reconstruction, thus reducing the useful muons and increasing the measurement time. It was found that the rejected PoCA events are ~30% of the total muons. This behavior is similar in all cases, independent of muon energy or material. Using the GMTE estimate, an increase in the useful muons will result in improved resolution or reduced measurement time. Figure 5 includes GMTE estimates when energy loss is taken into account. In this case, the muon path estimation appears to be slightly improved, as expected. The simulated muon path is within the 2σ and 3σ envelopes. An exception to this is the case in which large angle scattering occurs. However, it has been pointed out that particles undergoing large angle scattering can be eliminated using angular cuts [24].

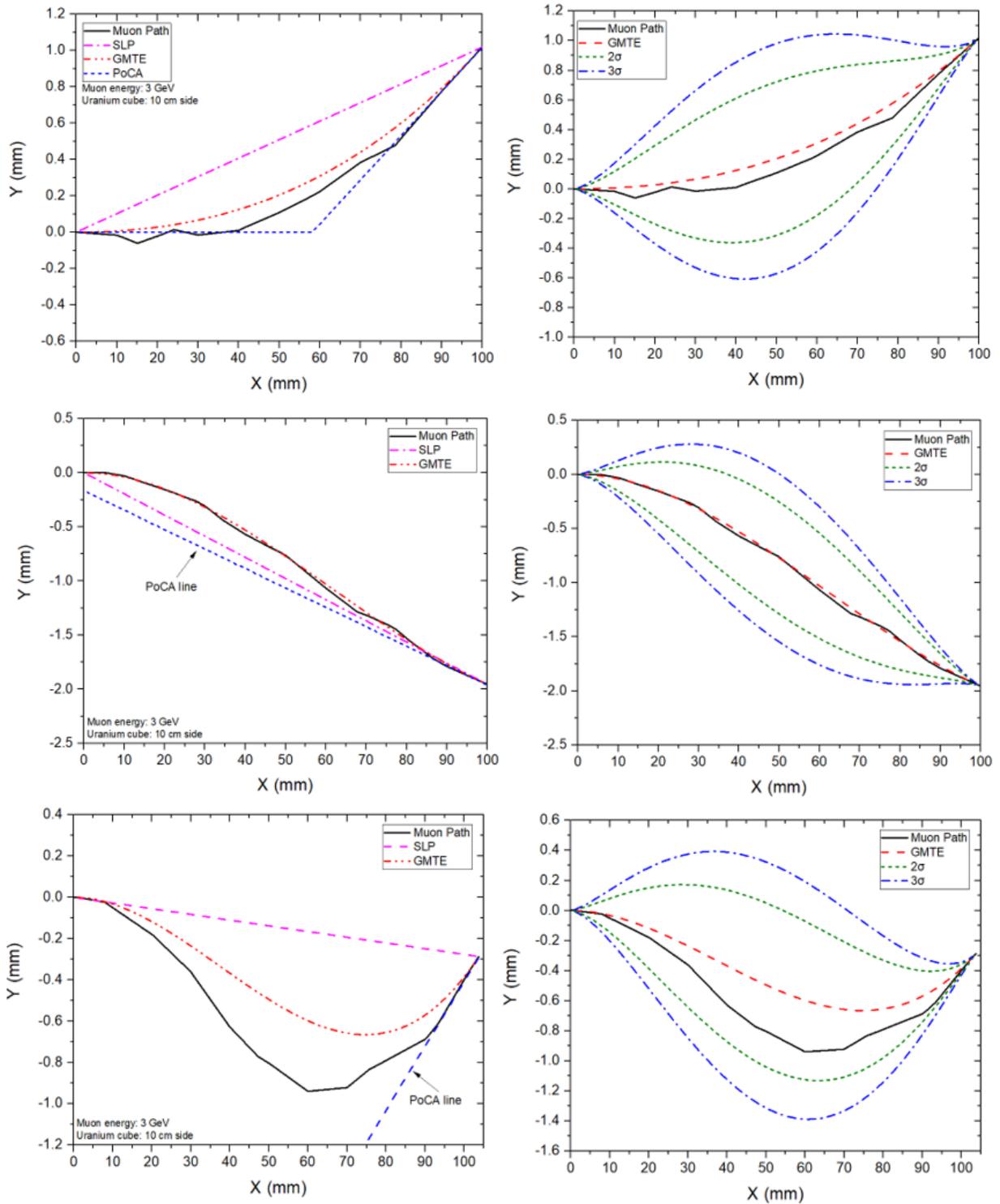

Figure 4. Examples of individual muon paths (projected in 2D) obtained from Geant4 (solid line), SLP, PoCA and MLP (left side) and the associated 2σ and 3σ envelopes (right side). In the two bottom examples, the PoCA line is outside the reconstruction boundaries.

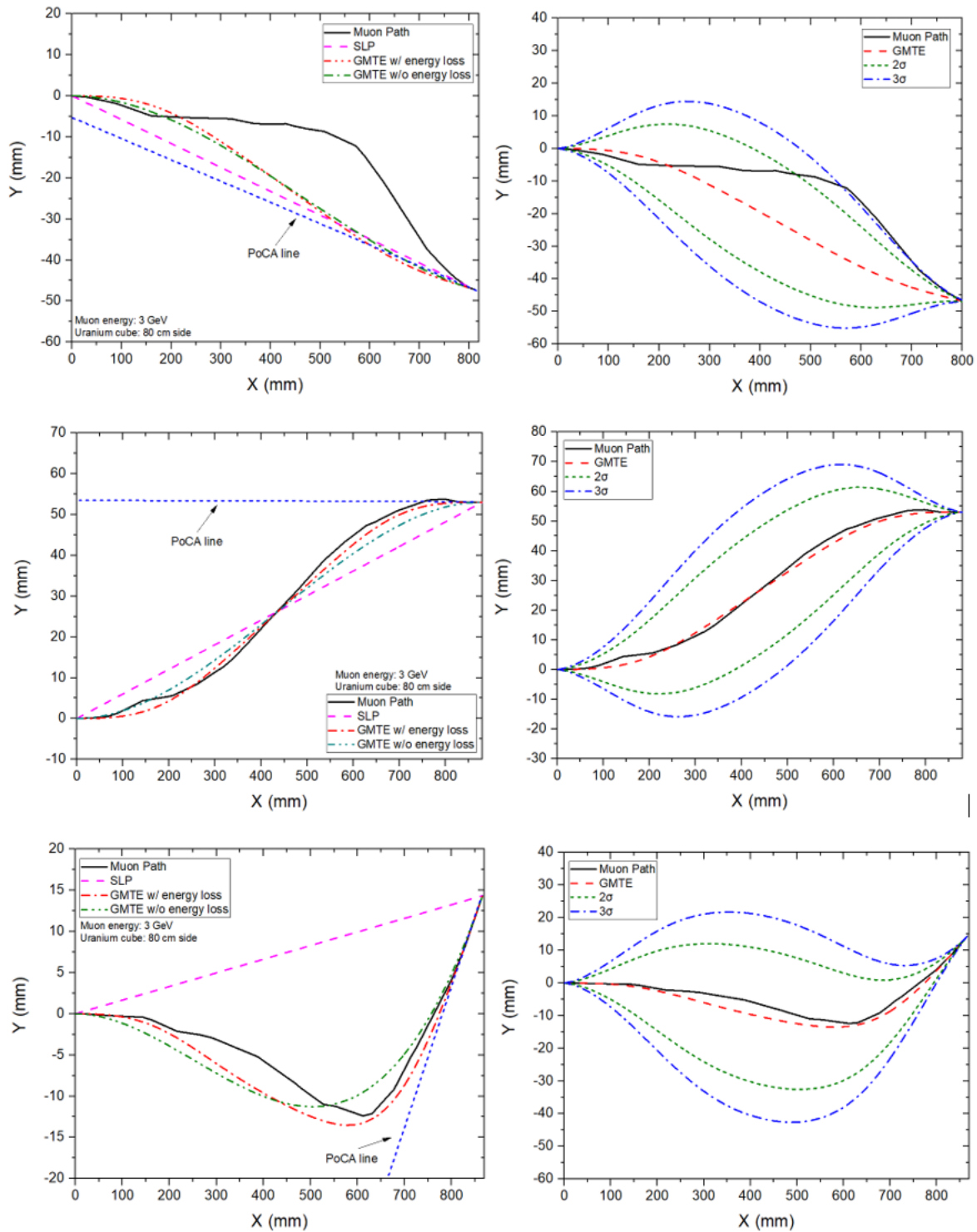

Figure 5. Examples of individual muon paths (projected in 2D) obtained from Geant4 (solid line), SLP, PoCA and MLP with (w/) and without (w/o) energy loss (left side) and the associated 2σ and 3σ envelopes (right side). In all examples, the PoCA line is outside the reconstruction boundaries. In the bottom-left example, a muon that underwent a large angle scattering collision that falls outside both error envelopes is shown.

A quantity of 100,000 muons were simulated, and the root mean square (RMS) deviation of the lateral displacement in the projection plane between the path estimates and the Geant4 simulated muon paths is shown in Figure 6. The largest RMS deviation occurs about halfway between entry and exit depth, where the paths are farthest from the known entry and exit points. The RMS deviation provided by the GMTE estimate is ~0.25 mm in the central region. The SLP and PoCA RMS deviations are ~0.45 mm and ~0.3 mm. This translates to a 45% and 16% improvement compared to SLP and PoCA, respectively. It is noted that in the PoCA RMS calculation, only the accepted PoCA events were considered. Using the PoCA events that occur outside the reconstructed boundaries would increase the RMS deviation considerably.

To compare the behavior of the GMTE for different muon energies and materials, the area under the curve (AUC) of the RMS error curve and the precision were used as figures of merit. The precision is defined as the maximum RMS error. The AUC and precision were calculated for different muon energies and materials. The results are shown in Figures 7 and 8. As expected, GMTE has a smaller overall AUC than either SLP or PoCA. In addition, the AUC decreases with increasing muon energy. This can be attributed to the smaller scattering variance that varies inversely to muon momentum. The AUC obtained for different materials is shown in Figure 8. The precision, or the peak value of the RMS curve, is also shown. High-Z materials tend to scatter muons more, so the expected error increases. However, the precision is on the order of 1.5 mm for 0.5 GeV, and it follows a decreasing trend with increasing muon energy.

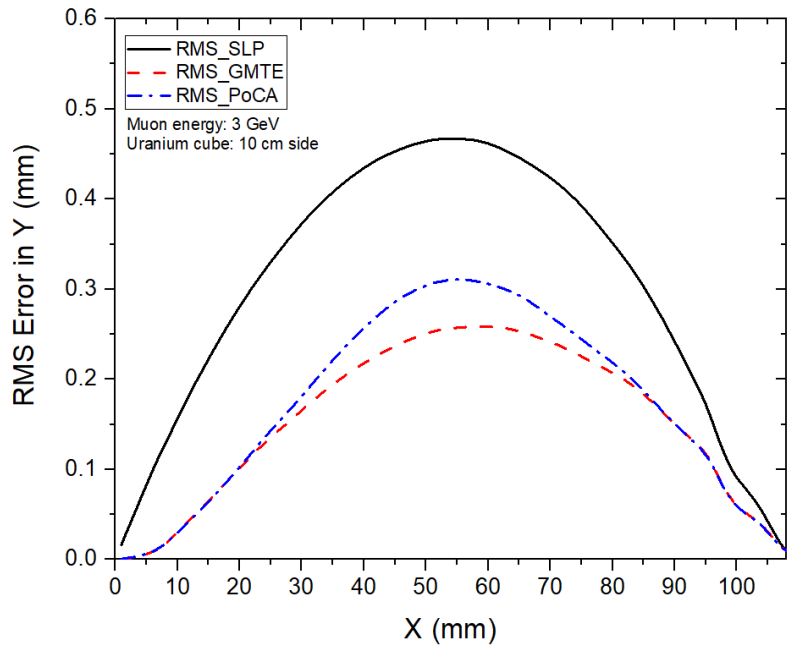

Figure 6. RMS deviation in lateral displacement between estimated path and Geant4 simulated path; curves show results for SLP (linear), GMTE and PoCA estimated paths.

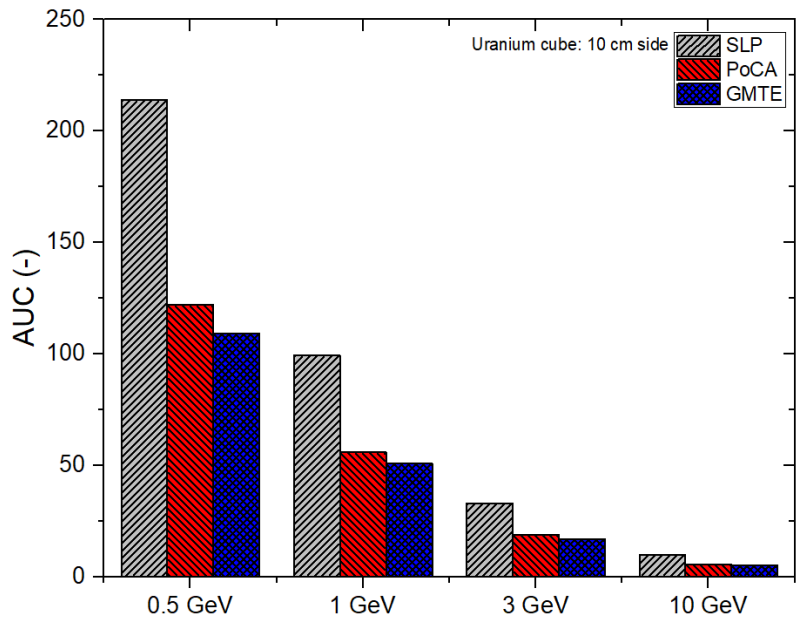

Figure 7. Area under curve (AUC) for different initial muon energies.

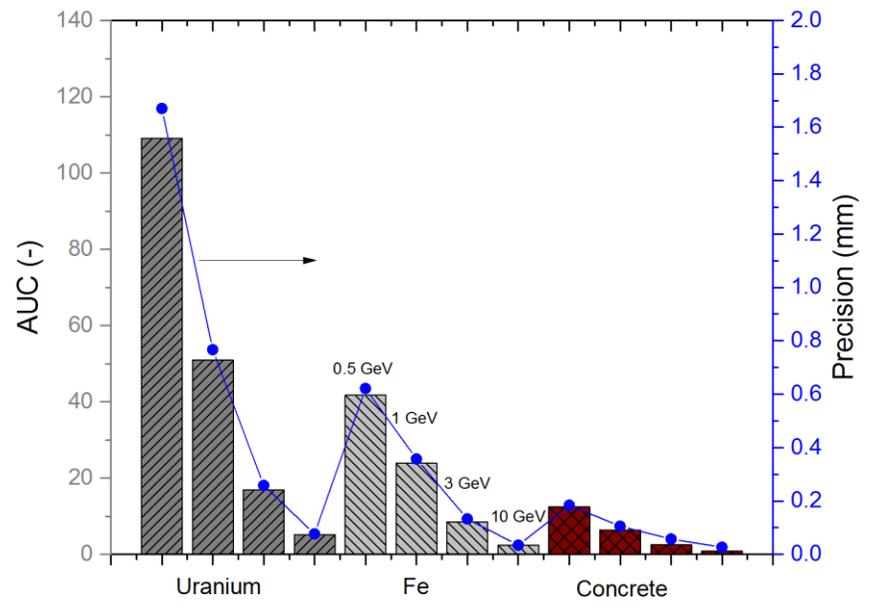

Figure 8. AUC and precision as a function of muon energy and material for GMTE (uranium cube, 10 cm side).

## 6. Conclusions

This paper explores the use of a physics-based algorithm for estimating the cosmic ray muon trajectory through a material when the incoming and outgoing positions and directions are known. This work presents an algorithm based on Bayesian theory, a Gaussian approximation of MCS, and generalized scattering and displacement moments to estimate the path of a cosmic ray muon in either uniform or nonuniform geometries. Further, the energy loss is calculated using the continuous slowing-down approximation and the minimum ionization property of muons to derive a linear relation. The algorithm's expected precision was assessed using Monte Carlo simulated muon trajectories for selected materials and geometries. The algorithm is expected to be able to predict muon tracks to less than 1.5 mm RMS for 0.5 GeV muons and 0.25 mm RMS for 3 GeV muons, a 15% and 50% improvement compared to PoCA and SLP algorithms, respectively. Further, a 30% increase in useful muon flux should be observed relative to PoCA. Muon track prediction improved for higher muon energies or smaller penetration depths where energy loss is not significant. The effect of energy loss due to ionization was investigated, and a linear energy loss relation that is easy to use was described. It is expected that the generalized muon trajectory estimation algorithm will produce muon tomographic images with improved quality and reduced noise, requiring fewer muons than the currently available muon reconstruction techniques. Future work will focus on further development of the GMTE and demonstration on complicated geometries such as cargo containers or dry nuclear fuel storage casks.

## Acknowledgments

This research was sponsored by the Laboratory Directed Research and Development Program of Oak Ridge National Laboratory, managed by UT-Battelle, LLC, for the US Department of Energy.

## References

[1] K. N. Borozdin et al., "Radiographic imaging with cosmic ray muons," *Nature*, Vol. 422, p. 277, 2003.

[2] V. Anghel et al., "A plastic scintillator-based muon tomography system with an integrated muon spectrometer," *Nucl. Instrum. Methods, Phys. Res. A*, vol. 798, pp. 12–23, Oct. 2015.

[3] S. Pesente et al., "First results on material identification and imaging with a large-volume muon tomography prototype," *Nucl. Instrum. Methods, Phys. Res. A*, vol. 604, no. 3, pp. 738–746, 2009.

[4] L. J. Schultz, Cosmic ray muon radiography, Thesis dissertation, Portland State University, 2003.

[5] P. Baesso, D. Cussans, C. Thomay, and J. Velthuis, "Toward a RPC based muon tomography system for cargo containers," *J. Instrum.*, vol. 9, p. C10041, Oct. 2014.

[6] S. Chatzidakis et al., "Interaction of Cosmic Ray Muons with Spent Nuclear Fuel Dry Casks and Determination of Lower Detection Limit," *Nucl. Instr. Methods Phys. Res. A*, Vol. 828, pp. 37–45, 2016.

[7] K. Nagamine, "Geo-tomographic Observation of Inner-structure of Volcano with Cosmic-ray Muons," *Journal of Geography* 104(7), pp 998–1007, 1995.

[8] C. J. Rhodes, "Muon tomography: looking inside dangerous places," *Science Progress*, 98(3), 2910299, 2015.

[9] A. Anastasio et al., "The MU-RAY experiment. An application of SiPM technology to the understanding of volcanic phenomena," *Nucl. Instrum. Methods, Phys. Res. A*, vol. 718, pp. 134–137, 2013.

[10] G.E. Hogan, et al., "Detection of High-Z Objects using Multiple Scattering of Cosmic Ray Muons," *AIP Conference Proceedings*, 698(1): p. 755–758, 2004.

[11] S. Riggi et al., *Muon tomography imaging algorithms for nuclear threat detection inside large volume containers with the Muon Portal detector*, arXiv:1307.0714v1, 2013.

[12] L. J. Schultz, et al., "Image reconstruction and material Z discrimination via cosmic ray muon radiography," *Nucl. Instrum. Methods, Phys. Res. A*, Section A (Accelerators, Spectrometers, Detectors and Associated Equipment), 519(3): p. 687, 2004.

[13] C. Jewett et al., "Simulation of the use of cosmic rays to image nuclear waste and verify the contents of spent fuel containers," *Proceedings of the Waste Management Conference*, Phoenix, AZ, 2011.

[14] P.M. Jeneson, "Large vessel imaging using cosmic ray muons," *Nucl. Instrum. Methods, Phys. Res. A,* Vol. 525, pp. 346–351, 2004.

[15] S. Chatzidakis, C. K. Choi, and L. H. Tsoukalas, "Interaction of cosmic ray muons with spent nuclear fuel dry casks and determination of lower detection limit," *Nucl. Instrum. Methods, Phys. Res. A*, vol. 828, pp. 37–45, Aug. 2016.

[16] C. Thomay, J. Velthuis, T. Poffley, P. Baesso, D. Cussans, and L. Frazão, "Passive 3D imaging of nuclear waste containers with muon scattering tomography," *J. Instrum.*, vol. 11, p. P03008, Mar. 2016.

[17] K.N. Borozdin, et al., "Cosmic Ray Radiography of the Damaged Cores of the Fukushima Reactors," *Physical Review Letters*, 109(15) 2012.

[18] M. Miyadera et al., "Imaging Fukushima Daiichi reactors with muons," *AIP Adv.*, vol. 3, 052133, 2013.

[19] L. J. Schultz et al., "Statistical reconstruction for cosmic ray muon tomography," *IEEE Transactions on Image Processing*, Vol. 16, No. 8, pp. 1985–1993, 2007.

[20] L.J. Schultz, et al., "Image reconstruction and material Z discrimination via cosmic ray muon radiography," *Nucl. Instrum. Methods, Phys. Res. A* (Accelerators, Spectrometers, Detectors and Associated Equipment), 519(3): p. 687, 2004.

[21] S. Chatzidakis et al., "Analysis of Spent Nuclear Fuel Imaging Using Multiple Coulomb Scattering of Cosmic Muons," *IEEE Trans. Nuc. Sci.*, vol. 63, 2866, 2016.

[22] D. Poulson et al., "Cosmic ray muon computed tomography of spent nuclear fuel in dry storage casks," *Nucl. Instr. Meth. Phys. Res. A*, vol. 842, 48–53, 2017

[23] Z. Liu et al., "Detection of Missing Assemblies and Estimation of the Scattering Densities in a VSC-24 Dry Storage Cask with Cosmic-Ray-Muon-Based Computed Tomography*,"* *J. Nucl. Mater. Manage.*, vol. 45, 12, 2017.

[24] Z. Liu et al., "Characteristics of Muon Computed Tomography of Used Fuel Casks Using Algebraic Reconstruction*,"* *IEEE Nucl. Sc. Symp. Conf. Record*, 2017.

[25] R.W. Shulte et al., "A maximum likelihood proton path formalism for application in proton computed tomography," *Med. Phys*. 35 (11), 2008.

[26] S. Agostinelli et al., "GEANT4 – A simulation toolkit," *Nucl. Instrum. Methods, Phys. Res. A*, 506, 250–303, 2003.

[27] S. Chatzidakis et al., "Exploring the Use of Muon Momentum for Detection of Nuclear Material Within Shielded Spent Nuclear Fuel Dry Casks," *Trans. Am. Nucl. Soc.*, Vol. 116, 190–193, 2017.

[28] J. M. Durham et al., "Cosmic ray muon imaging of spent nuclear fuel in dry storage casks," *J. Nucl. Mater. Manage.*, vol. 44, p. 3, 2016.

[29] G. Molière, "Theorie der Streuung schneller geladenen Teilchen I Einzelstreuung am abgeschirmten Coulomb-Feld," Zeitschrift Für Naturforsch, 133–145, 1947.

[30] H. A. Bethe, "Moliere's theory of multiple scattering," *Physical Review*, 89, 1256, 1953.

[31] K. Hagiwara, "Review of particle physics," *Phys. Rev. D*, vol. 66, no. 1, p. 010001, 2002.

[32] A. A. M. Mustafa & D. F. Jackson, "Small-angle multiple scattering and spatial resolution in charged particle tomography," *Phys. Med. Biol.*, 26, 461–472, 1981.

[33] V. L. Highland, "Some practical remarks on multiple scattering," *Nucl. Inst. Meth.*, 129, 497–199, 1975.

[34] G. R. Lynch & O. I. Dahl, "Approximations to multiple Coulomb scattering, "*Nucl. Instrum. Methods, Phys. Res. B*, 58, 6–10, 1991.